\documentclass[twocolumn,superscriptaddress,amsmath,amssymb,floatfix]{revtex4}
\usepackage{amssymb}
\usepackage{amsmath}
\usepackage{color}
\usepackage[pdftex]{graphicx}

\definecolor{jerem}{rgb}{1, 0, 0}
\definecolor{alex}{rgb}{0, 0, 1}
\definecolor{raph}{rgb}{0, 1, 0}
\definecolor{valide}{rgb}{0, 1, 1}
\definecolor{olivier}{rgb}{0.125, 0.26, 0.07}

\usepackage{mathtools}

\usepackage{enumerate}
\usepackage{mathtools}
\usepackage{stmaryrd}

\usepackage{enumitem}

\usepackage{textcomp}
\usepackage{gensymb}
\def \equi#1{\mathrel{\mathop{\kern 0pt\sim}\limits_{#1}}}

\begin{document}
\title{Joint statistics of  space and time exploration of $1d$ random walks}

\author{J. Klinger}
\affiliation{Laboratoire de Physique Th\'eorique de la Mati\`ere Condens\'ee, CNRS/Sorbonne Université, 
 4 Place Jussieu, 75005 Paris, France}
\affiliation{Laboratoire Jean Perrin, CNRS/Sorbonne Université, 
 4 Place Jussieu, 75005 Paris, France}
\author{A. Barbier-Chebbah}
\affiliation{Laboratoire de Physique Th\'eorique de la Mati\`ere Condens\'ee, CNRS/Sorbonne Université, 
 4 Place Jussieu, 75005 Paris, France} 
\author{R. Voituriez}
\affiliation{Laboratoire de Physique Th\'eorique de la Mati\`ere Condens\'ee, CNRS/Sorbonne Université, 
 4 Place Jussieu, 75005 Paris, France}
\affiliation{Laboratoire Jean Perrin, CNRS/Sorbonne Université, 
 4 Place Jussieu, 75005 Paris, France}
 \author{O.B\'enichou}
\affiliation{Laboratoire de Physique Th\'eorique de la Mati\`ere Condens\'ee, CNRS/Sorbonne Université, 
 4 Place Jussieu, 75005 Paris, France}
 

\begin{abstract}
The statistics of  first-passage times of random walks  to  target sites   has proved to play a key role in determining the {\it kinetics} of space exploration in various contexts . In parallel, the number of distinct sites visited by a random walker and related observables have been introduced to characterize the {\it geometry} of space exploration. 
 Here, we address the question of the {\it joint} distribution of   the first-passage time to a target and the number of distinct sites visited when the target is reached, which fully quantifies the coupling between kinetics and geometry  of search trajectories. Focusing on  1-dimensional systems, we present  a general method and  derive explicit expressions of this joint distribution for several representative examples of Markovian search processes. In addition, we  obtain a general scaling form, which  holds also for non Markovian processes and captures the general dependence of the joint distribution on its space and time variables. We argue that the joint distribution has important applications to various problems, such as a  conditional form of the Rosenstock trapping model, and the persistence properties of  self-interacting random walks.

\end{abstract}

\maketitle

Quantifying the efficiency of space exploration by random walkers is a key issue involved in a variety of situations. Applications range from reactive particles diffusing in the presence of catalytic sites, 
  living organisms looking for  resources, to  robots  cleaning or demining a given area \cite{Hughes:1995,Benichou:2011fk,Viswanathan:2008}. In this context, two important  classes of observables have been considered.  
 

First, the statistics of  first-passage times (FPTs)  to  target sites of interest  has proved to play a key role in determining the {\it kinetics} of space exploration  \cite{Redner:2001a,bookSid2014,Bray:2013}. The case of first-passage times  in  confined domains  was found to be particularly relevant to assess the efficiency of target  search processes, and has lead to an important activity  \cite{Condamin2007,Benichou:2014fk, Cheviakov:2010,Schuss2007}. Related observables, such as the cover time of a domain  \cite{Brummelhuis:1991ys,Brummelhuis:1992,Chupeau:2015sf} or the occupation time of a sub domain  have also been considered in this context \cite{Benichou:2005a,Weiss:1996,Burov:2007,Condamin:2008}.

A second class of observables has been introduced to characterize the {\it geometry} of the territory explored by random walkers. In particular,    the number of distinct sites visited  (or the so called Wiener sausage in a continuous setting) by a random walker after $n$ step, which quantifies the overall territory swept by the random walker,  has been the focus of many studies  with a broad range of applications \cite{Hughes:1995,Weiss:1994,Larralde:1992a,Berezhkovskii:1989}. 
 Notable extensions include  the 
number of distinct sites visited by $p$ independent walkers \cite{Larralde:1992a},   the case of  fractal geometries \cite{Blumen:1986,D.Ben-Avraham:2000}, or the case of random stopping times  \cite{Dayan:1992,Kearney:2005wo,Krapivsky:2010wt,Klinger:2021aa}.

Even if  it is clear that both classes of observables are coupled, so far  kinetic and geometric properties of exploration have been mainly discussed  independently, with the notable exception of \cite{Randon-Furling:2007vu}. Qualitatively,  the first-passage time to a target of a generic stochastic process carries information about the territory visited before hitting the target : large values of the first-passage time imply large values of the visited territory. However, the quantitative determination of this coupling is still lacking.

Here, we address the question of the {\it joint} distribution of  the first-passage time to a target and the number of distinct sites visited when the target is reached, which fully quantifies this coupling and  gives access to a refined characterization of search trajectories. To the best of our knowledge, this quantity has never been studied so far. The joint law provides two conditional distributions, which allow to answer quantitatively the following questions : ($Q_1$) What is the territory visited by a random walker knowing that it reached a target (and stopped or exited the domain)  after a given time?  ($Q_2$) How long does it take a random walker to reach a target knowing that it has visited a given number of distinct sites before? We anticipate that these quantities could have applications in various situations where only partial  information -- either kinetic or geometric -- on trajectories is accessible.    

\begin{figure}
	\centering
	\includegraphics[scale=0.25]{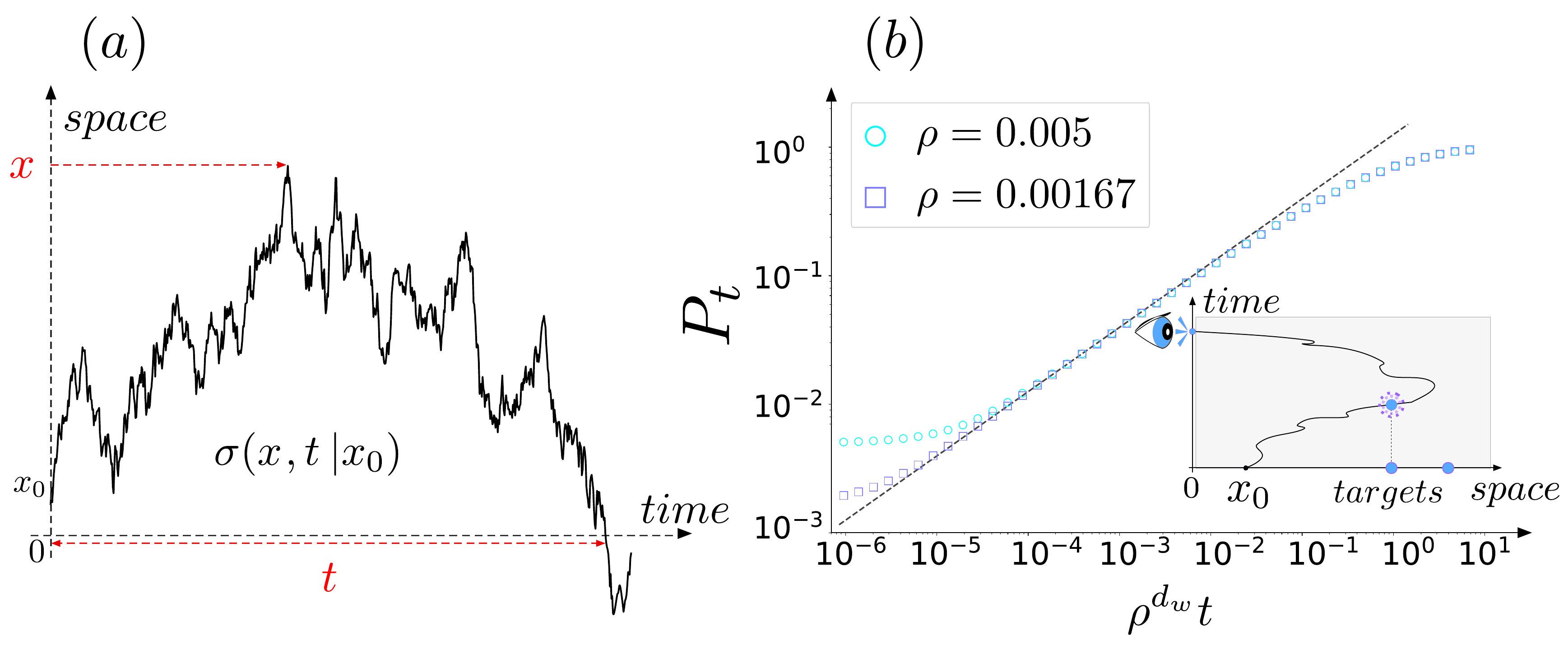}
	\caption{(a) Starting from $x_0$, the random walker crosses 0 for the first time at time $t$, having explored up to a distance $x$ from the origin. The joint law $\sigma(x,t|x_0)$ is the density probability function of such joint events. (b) Consider a random searcher evolving in a (greyed out)  domain, filled with Poisson distributed targets. Having as only information the time $t$ of exit from the domain, we display the probability $P_t$ of encounter with at least one target in terms of the rescaled variable $\rho^{d_w}t$ in the case of a Brownian searcher of diffusion coefficient $D=1/2$. Numerical integration of the exact result \eqref{eq : survival_prob} (symbols) and asymptotic scaling form \eqref{scalingRS} (dashed line) are shown.}
	\label{fig : panel_1}
\end{figure}

{\it Summary of the results.} We tackle this general question in the case of 1-dimensional processes, and determine  the joint distribution $\sigma(s,n|s_0)$ of the FPT $n$ at the target site $0$ and the  number  $s$ of distinct sites visited by a random walker starting from $s_0$ (see Fig. \ref{fig : panel_1}(a), where $x,t$ are the continuous counterparts of $s,n$)\footnote{This should not be confused with  the  joint distribution of  the maximum and the time for reaching the maximum of a Brownian motion   derived in \cite{Randon-Furling:2007vu}}.  Our approach applies to    general (space and time) discrete or continuous   random walkers,  evolving in a semi infinite or  finite domain, and yields   fully explicit expressions of $\sigma(s,n|s_0)$ for several representative examples of Markovian  processes, such as  simple symmetric and biased random walks, persistent random walks \cite{Hughes:1995,Weiss:1994} or resetting random walks \cite{Evans:2011,Evans:2020aa}, whose definitions are recalled below. In addition, we  derive a general scaling form of $\sigma(s,n|s_0)$ in the large $s,n$ regime, which  holds also for non Markovian processes and captures the general dependence on $s_0,s,n$. Several applications of these central results are then discussed. First, we determine the efficiency of a schematic catalytic reaction \cite{Rosenstock:1961} by deriving the probability that a diffusing  particle has reacted in a domain with Poisson distributed targets before exiting the domain, knowing the exit time (see Fig. \ref{fig : panel_1}(b)). 
Second, we show that the knowledge of the joint distribution $\sigma(s,n|s_0)$ for simple random walks  is actually required for determining  first-passage properties of a class of strongly  non Markovian processes, namely self-interacting random walks \cite{amit_asymptotic_1983,pemantle_survey_2007,grassberger_self-trapping_2017,foster_reinforced_2009,stevens_aggregation_1997}. More precisely, with the help of the  joint distribution, we derive exactly  the full large time behavior of the FPT density of   the so-called self-attracting walk (SATW) \cite{sapozhnikov_self-attracting_1994}, which has been studied in the context of random search processes as a prototypical example of processes with long-range memory \cite{boyer_modelling_2010, boyer_non-random_2012, borger_are_2008,PhysRevLett.119.140603} and has important  applications in the theoretical description of the trajectories of living organisms such as cells  \cite{dAlessandro:2021vx}.

{\it Discrete processes.} We  first consider the case of a general Markovian discrete (in space and time) process,  which leaves no holes in its trajectory; in other words, the set of visited sites is assumed to be at all times the finite range $\llbracket s_{\rm min},s_{\rm max}\rrbracket$ defined by the min ($ s_{\rm min}$) and max ($ s_{\rm max}$) values of the random walker's positions. This last hypothesis will hold for all processes presented in what follows. Denoting by $s_0>0$ the starting site, $n$ the step at which the walker reaches the target $0$ for the first time and $s$ the number of distinct  sites visited up to this random stopping time, we derive a systematic procedure to obtain the joint law $\sigma(s,n|s_0)$. In turn, this  joint law gives immediate access to the conditional probabilities mentioned above, \textit{ie} (i) the distribution  of the number $s$ of distinct sites visited before reaching 0  knowing that the random walker has reached 0  at step $n$ :
\begin{equation}\label{Gs}
G_{sp}(s|n,s_0)=\frac{\sigma(s,n|s_0)}{\sum_{s'=s_0}^{\infty}\sigma(s',n|s_0)}\equiv \frac{\sigma(s,n|s_0)}{F_{\underline{0}}(n|s_0)},
\end{equation}
where $F_{\underline{0}}(n|s_0)$ is the usual FPT distribution to 0 and (ii) the distribution of the FPT to 0  knowing that  $s$ distinct sites have been visited before reaching $0$ :
\begin{equation}\label{Gt}
G_{tm}(n|s,s_0)=\frac{\sigma(s,n|s_0)}{\sum_{n'=0}^{\infty}\sigma(s,n'|s_0)}\equiv \frac{\sigma(s,n|s_0)}{\mu_0(s|s_0)},
\end{equation}
where $\mu_0(s|s_0)$ is the distribution of the maximum $s$ before reaching 0.

Let us  denote $F_{\underline{0},s}(n|s_0)$ the probability that the walker reaches zero for the first time at step $n$, without  ever reaching $s$,  and make a partition over the rightmost site $s'$ visited before reaching zero. Because the walker  reaches 0 before $s$, one necessarily has $s'\in\llbracket s_0,s-1 \rrbracket$, which yields $F_{\underline{0},s}(n|s_0)=\sum_{s'=s_0}^{s-1}\sigma(s',n|s_0)$.
Note that this relation still holds for non Markovian processes.
Equivalently, we obtain the key relation 
\begin{equation}\label{eq : build_block}
\sigma(s,n|s_0)=F_{\underline{0},s+1}(n|s_0)-F_{\underline{0},s}(n|s_0)\equiv D_s F_{\underline{0},s}(n|s_0),
\end{equation}
which  allows one to write the joint law $\sigma$ explicitly in terms of the quantity $F_{\underline{0},s}(n|s_0)$.

We next provide  a procedure based on backward equations  to  derive the probability  $F_{\underline{0},s}(n|s_0)$ in presence of two absorbing sites $0$ and $s$ for a given Markovian stochastic process. 
 In this case, the propagator $P(s,n|s_0)$, \textit{ie} the probability for the walker to be at site $s$ after $n$ steps, obeys the backward equation $P(s,n+1|s_0)=\mathcal{L}_{s_0}\left[P(s,n|s_0)\right]$ \cite{Hughes:1995,Kampen:1992a}, obtained by partitioning over the first step of the walk, 
where $\mathcal{L}_{s_0}$ is a  linear operator acting on $s_0$.
For instance, in the case of a simple  random walk, $\mathcal{L}_{s_0}\left[P(s,n|s_0)\right]=\frac{1}{2}P(s,n|s_0+1)+\frac{1}{2}P(s,n|s_0-1)$. It is easily seen  that $F_{\underline{0},s}(n|s_0)$ obeys the same backward equation for $0<s_0<s$ and, introducing the generating function  $\tilde{F}_{\underline{0},s}(\xi|s_0)=\sum_{n=0}^{\infty} \xi^n F_{\underline{0},s}(n|s_0)$,  we obtain:
\begin{equation}\label{eq : masterF}
\tilde{F}_{\underline{0},s}(\xi|s_0)=\xi\mathcal{L}_{s_0}\left[\tilde{F}_{\underline{0},s}(\xi|s_0)\right].
\end{equation}
Reminding that both $0$ and $s$ are absorbing boundaries, we have that,  for any $n>0$, $F_{\underline{0},s}(n|0 \text{ or } s)=0$ whereas $F_{\underline{0},s}(0|0)=1$ and $F_{\underline{0},s}(0|s)=0$. In terms of generating functions, we obtain the following boundary conditions:  
\begin{equation}\label{eq : BC}
\tilde{F}_{\underline{0},s}(\xi|0)=1 \ ;\ 
\tilde{F}_{\underline{0},s}(\xi|s)=0.
\end{equation}
Eq. \eqref{eq : masterF}, completed by \eqref{eq : BC}, fully determines  $\tilde{F}_{\underline{0},s}(\xi|s_0)$. Making use of \eqref{eq : build_block}, we then derive the generating function of the joint law $\sigma$.

As an illustration, we obtain in the case of a simple random walk (see supplementary material (SM))
\begin{equation}\label{sigmadiscret}
\tilde{\sigma}(s,\xi|s_0)=\frac{r_+-r_-}{r_+^{s}-r_-^{s}}\frac{r_+^{s_0}-r_-^{s_0}}{r_+^{s+1}-r_-^{s+1}}
\end{equation}
where $r_{\pm}=\frac{1}{\xi}(1\pm \sqrt{1-\xi^2})$. Further illustration is provided in SM, where explicit expressions of  $\tilde{\sigma}(s,\xi|s_0)$ are determined for the important examples of biased random walks (for which a step is taken to the right  with probability $p$, and to the left with probability $1-p$), persistent random walks (for which each step is taken identical to the previous one with probability $p$) \cite{Ernst:1988fk,Tejedor:2012ly} and resetting random walks (for which at each step the walker has a probability $\lambda$ to jump back to its initial position) \cite{Evans:2011,Evans:2020aa,PhysRevE.92.052126,PhysRevLett.113.220602}. Finally, in each case, a  series expansion with respect to $\xi$  gives  access to an exact determination of   $\sigma(s,n|s_0)$ (see SM for validation by  numerical simulations), which constitutes the main result of this section; its physical implications are commented below (see {\it discussion} and {\it applications}).

{\it Continuous space and time.} This method is easily adapted to continuous space and time $(x,t)$ Markovian processes. Defining $F_{\underline{0},x}(t|x_0)$ as the probability density to reach 0 before $x$ at time $t$,
the continuous counterpart of Eq. \eqref{eq : build_block} reads:
\begin{equation}\label{eq : continuous_jointlaw}
\sigma(x,t|x_0)= D_x F_{\underline{0},x}(t|x_0)
\end{equation}
where  here $D_x$ is the differential operator with respect to $x$, 
and the Laplace transform $\tilde{F}_{\underline{0},x}(p|x_0)=\int_0^\infty e^{-pt} F_{\underline{0},x}(t|x_0){\rm d}t$ satisfies the continuous counterpart of Eq \eqref{eq : masterF}, \eqref{eq : BC} (see SM). 
As an explicit example,  for Brownian  diffusion with diffusion coefficient $D$, it is found that the joint law is given by
\begin{equation}\label{eq : laplace_invert}
\begin{split}
\sigma(x,t|x_0)=\frac{2D\pi}{x^3}\sum_{k=1}^{\infty}&e^{-(k\pi)^2D\tau}k\sin(k\pi\Tilde{x}_0) \times\\
&\times\left[2( k\pi)^2D\tau-2-\frac{k\pi\Tilde{x_0}}{\tan(k\pi\Tilde{x}_0)}\right],
\end{split}
\end{equation}
where $\tilde{x_0}=\frac{x_0}{x}$ and $\tau=\frac{t}{x^2}$. 
Explicit expressions of $\tilde{\sigma}$ for other continuous Markov processes (biased diffusion and continuous resetting) are presented in SM. Importantly, it is also shown in SM that our approach can be further extended to the case of  continuous space but discrete time processes, also known as jump processes, as well as Markovian processes in confined domains.

{\it General scaling form.} Beyond the case of Markovian processes, we now show that the joint law $\sigma$  assumes a general scaling form for  symmetric processes, which  holds even in the non Markovian case and elucidates its dependence on the parameters $s,s_0,n$. Because we are interested only  in the large time and space limit, we adopt a continuous formalism and make use of the variables $x,x_0,t$.  Extending an approach given in \cite{Majumdar:2010,Levernier:2018qf},  we derive below a general scaling form for $F_{\underline{0},x}(t|x_0)$, which leads
to the asymptotic behavior of $\sigma(x,t,x_0)$. 

First, note that walkers reaching $x$ before 0 do not contribute to the probability $F_{\underline{0},x}(t|x_0)$. Hence, for times shorter than the typical time $T_{typ}\propto x^{d_w} $ needed to reach $x$ (which defines the walk dimension $d_w$ of the process),  $F_{\underline{0},x}(t|x_0)$ behaves as the first-passage time density $F_{\underline{0}}(t|x_0)$ in a semi-infinite domain, with a {\it single} target in 0. We now assume that this quantity has an algebraic decay with time for $t\to\infty$, quantified by the persistence exponent $\theta$ of the process: $F_{\underline{0}}(t|x_0)\sim k(x_0) t^{-(\theta+1)}$, where $k(x_0)\propto x_0^{d_w\theta}$ for $x_0 \gg 1$ \cite{Bray:2013}. Because almost all random walkers have either reached $0$ or $x$ at times $t \gg x^{d_w}$, we write 
\begin{equation} 
F_{\underline{0},x}(t|x_0)\sim F_{\underline{0}}(t|x_0)g(\frac{t}{x^{d_w}})\sim k(x_0) t^{-(\theta+1)}g(\frac{t}{x^{d_w}})
\end{equation}
where $g$ is  a smooth cut-off function  with  $g(0)=1$ and $g(y)$ vanishes for large $y$. Finally,
 with the help of \eqref{eq : continuous_jointlaw}, we obtain the general scaling form for the joint law in the scaling limit defined by $x\to\infty,t\to\infty$ with $\tau=t/x^{d_w}$ fixed  :
\begin{equation}\label{eq : scaling_form}
\sigma(x,t|x_0)\sim\frac{h(x_0)}{x^{d_w(\theta+1)+1}}f(\tau)
\end{equation}
where, defining $f_1(\tau)=-d_wg'(\tau)\tau^{-\theta}$ and ${\cal N}=\int_0^\infty f_1(\tau) d\tau$, we have $h(x_0)= k(x_0){\cal N}$ and $f=f_1/{\cal N}$. In addition,  $h(x_0)\propto x_0^{d_w\theta}$ for $x_0 \gg 1$, and $f(\tau)$ is a normalized process dependent function.

Of note, integrating equation \eqref{eq : scaling_form} over $t$ recovers the distribution of the maximum  $\mu_0(x|x_0)=h(x_0)x^{-(d_w\theta+1)}$ before reaching 0, 
in agreement with known results \cite{Majumdar:2010}. In turn,  this provides a simple physical interpretation of $f(\tau)$. Making use of \eqref{Gt}, we obtain the  conditional density $G_{tm}(t|x,x_0)\sim\frac{1}{x^{d_w}}f(\tau)$. Thus, $f(\tau)$ is the density of the rescaled variable $\tau$ conditioned by the value of the maximum $x$. In particular, we stress that $f$ is  independent of $x_0$.

\begin{figure}
\centering
\includegraphics[scale=0.24]{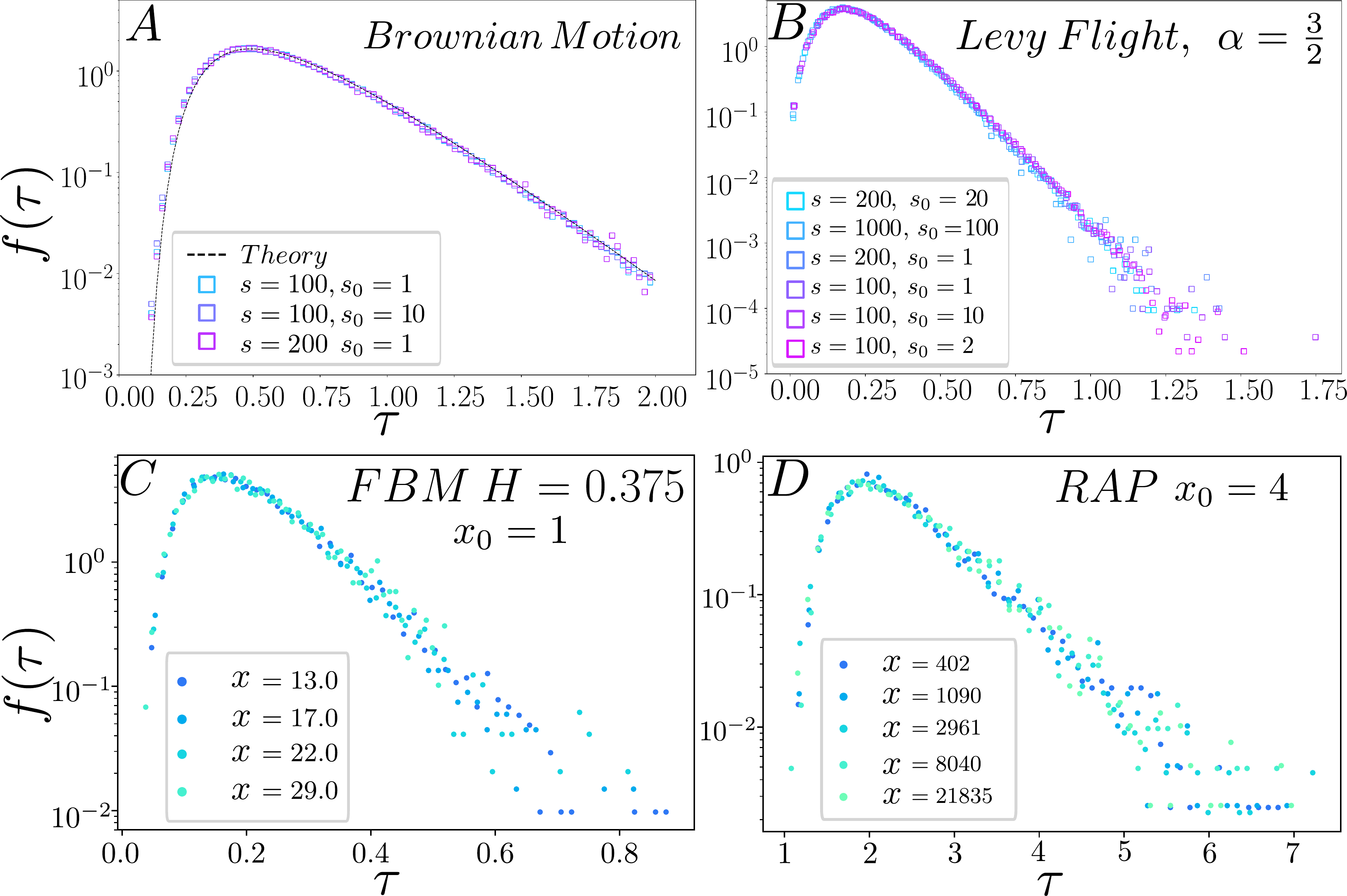}
\caption{Conditional distribution $f(\tau)$ of the rescaled variable $\tau$ (see text). Distributions are drawn for fixed $s$ (discrete space) or $x$ (continuous space) and collapse. \textbf{A,B} - Markovian Processes ; \textbf{C,D} Non Markovian Processes. See SM for details on simulations.}
\label{fig : continuous_sim}
\end{figure}

The general relation \eqref{eq : scaling_form} is confirmed in Fig \ref{fig : continuous_sim} by numerical simulations for  representative examples of  both Markovian processes (simple random walks  and Riemann walks, \textit{ie} discrete space and time Levy flights \cite{Hughes:1995}),  and non Markovian processes (Fractional Brownian Motion \cite{MANDELBROT1968} and the Random Acceleration Process \cite{Bicout:2000}, see SM for definitions).
Indeed, we find that the conditional density of the FPT knowing the territory covered,  which a priori depends on the  variables $t,\ x,\ x_0$, can in fact be rewritten as the distribution $f(\tau)$ of the single reduced variable $\tau$, as shown by the data collapse in the figure. 
Next, thanks to the exact Eq. \eqref{Gt}, and the exact  scaling of the distribution $\mu_0$ of the maximum reminded above \cite{Majumdar:2010},  this observed scaling of $f$ directly confirms \eqref{eq : scaling_form}.

In the case of diffusive random walks, $f(\tau)$ can be determined explicitly by taking $x\rightarrow \infty$ and $t\rightarrow \infty$ with $\tau$ fixed in Eq. \eqref{eq : laplace_invert}: 
\begin{equation}\label{eq : scaling_discrete_diffusion}
    f_{BM}(\tau)=2D\pi^2\sum_{k=1}^{\infty}e^{-(k\pi)^2D\tau}k^2\left[2( k\pi)^2D\tau-3\right].
\end{equation}
Of note, this asymptotic conditional distribution holds for any symmetric Markovian random walk satisfying the central limit theorem. 


Similarly (see SM), the other conditional distribution defined in \eqref{Gs} can be written from \eqref{eq : scaling_form} as $G_{sp}(x|t,x_0)\sim \frac{1}{t^{1/d_w}}\phi(\chi)$ where the density of the rescaled variable $\chi=x/t^{1/d_w}$ is given in terms of $f$ by :
\begin{equation}\label{phi}
   \phi(\chi)=\frac{\chi^{-d_w(\theta+1)-1}f(\chi^{-d_w})}{\int_0^\infty u^{-d_w(\theta+1)-1}f(u^{-d_w}) du}.
\end{equation}
The agreement of this result  with numerical simulations is shown in SM.

{\it Discussion}. The above results yield both exact expressions 
 of the joint law for Markovian processes, and  scaling forms for general non Markovian processes, and have important implications. (i) The joint law, because it gives access to all correlation functions $\langle x^n t^m\rangle$, fully quantifies the coupling between the kinetics of space exploration and the territory explored by a random walker. This coupling manifests itself in the dependence of $\sigma$ on the rescaled variable $\tau=t/x^{d_w}$. (ii) The joint law yields the conditional distributions $G_{sp}$ (see \eqref{Gs}) and $G_{tm}$ (see \eqref{Gt}), which provide new insights in the quantification of space exploration, and  in particular explicit answers to the questions $Q_1,Q_2$ raised in  introduction.   Below, we further illustrate the importance of the joint law and  turn to examples of applications of our results.

{\it Application -- Conditional Rosenstock problem}.  The above results provide as a by product an explicit solution to a conditional version of the celebrated  Rosenstock problem \cite{Rosenstock:1961,Hughes:1995}. We consider 
a reactive diffusing  particle that enters a 1-dimensional chemical reactor at $x_0$ and leaves it at 0. The reactor  contains  Poisson distributed catalytic point-like sites of density $\rho$, which trigger a reaction  upon encounter with the reactive particle (see Fig.\ref{fig : panel_1} (b)). 
The efficiency of such  schematic catalytic reaction can be quantified by  the probability $P_t$ that the  reactive   particle has reacted with a catalytic site  before exiting the domain, knowing the exit time $t$. This is readily obtained as
\begin{equation}\label{eq : survival_prob}
    P_t=\int_0^\infty (1-e^{-\rho x})G_{sp}(x|t,x_0)\text{d}x.
\end{equation}
The determination of $P_t$  thus requires $G_{sp}$, and therefore the joint law. Making use of the general scaling \eqref{eq : scaling_form}, we obtain the large time scaling behaviour : 
\begin{equation}\label{Ptscaling}
    P_t\equi{t\to\infty}\int_0^\infty (1-e^{-\rho t^{1/d_w}u})\phi(u)\text{d}u;
\end{equation} 
this shows that $P_t$ is asymptotically a function of the reduced variable $\rho t^{1/d_w}$ only, with $P_t\propto \rho t^{1/d_w}$ for $ \rho t^{1/d_w}\to 0$. Equation \eqref{Ptscaling} provides, thanks to \eqref{phi}, an explicit determination of $P_t$ for all processes for which $\sigma$ (and thus $f$)  is known, and in particular elucidates its dependence on 
 the exit time $t$ from the domain (see Fig.\ref{fig : panel_1}(b)). On the example of Brownian motion, one obtains (for $ x_0^2 /D\ll t\ll 1/(D\rho^2)$):
\begin{equation}\label{scalingRS}
    P_t\sim\sqrt{\pi}\rho (Dt)^{\frac{1}{2}}.
\end{equation}

{\it Application --Self-interacting walkers.} Next, we show that  the joint law can be needed to obtain the  first-passage time distribution. This is  the case of self-interacting random walks, which are defined generically as random walks whose jump probabilities at time $n$ depend on the full set of  visited sites at earlier times $n'<n$. We focus on the example of the $1d$ self-attracting walk (SATW) \cite{sapozhnikov_self-attracting_1994}, which has been studied in the context of random search processes as  a prototypical example of process with long-range memory, and has recently proved to be relevant to describe the dynamics of motile cells  \cite{dAlessandro:2021vx}.
At each time step, if both its neighboring sites have already been visited, the random walker hops on either of them with probability $1/2$. However, if one of them has never been visited, it is chosen with probability $\beta$. Note that this can either be an attractive effect ($\beta<1/2$) or a repulsive one ($\beta>1/2$). Since the dynamics of the walk is completely determined by the location of unvisited  sites, the determination of the first-passage time distribution requires the knowledge of all times at which 
unvisited  sites have been  discovered. 
Denoting here $F_{0,\underline{s}}(n|s_0)$ the probability to reach $s$ before $0$ for the first time at step $n$, knowing that the sites $\llbracket1,s-1\rrbracket$ have already been visited, the generating function of $\sigma(s,n|1) $ can be written as :
\begin{equation}\label{eq : convolution}
\tilde{\sigma}(s,\xi|1)=\frac{\xi}{2}\left(\prod_{s'=3}^s \tilde{F}_{0,\underline{s'}}(\xi|s'-1)\right)\tilde{F}_{\underline{0},s+1}(\xi|s) 
\end{equation}
Solving for $\tilde{F}_{0,\underline{s}}(\xi|s_0)$ yields an explicit expression of $\Tilde{\sigma}$ (see SM). For large $s$ and $n$, with $\tau=\frac{n}{s^2}$ fixed, this yields  $\sigma(s,n| s_0)= h(s_0) s^{-\frac{1-\beta}{\beta}-3}f_{SATW}(\tau)$ where $h(1)=\frac{\Gamma(-2 + 2/\beta)}{\Gamma(-1 + 1/\beta)} \frac{(1-\beta)}{\beta}$ and $h(s_0) \propto s_0^{\frac{1-\beta}{\beta}}$ for large $s_0$ \footnote{Since for the SATW $d_w=2$  and $\theta=\frac{1-\beta}{2\beta}$ \cite{BarbierChebbah:2020aa} the joint law obeys the general scaling form \eqref{eq : scaling_form}}. Finally, the conditional distribution $f_{SATW}$ is defined by its strikingly simple Laplace transform:
\begin{equation}\label{eq : conditional2}
\tilde{f}_{SATW}(p) = \int_0^\infty e^{-pu}f_{SATW}(u){\rm d}u =  \left(\frac{\sqrt{2p}}{\sinh(\sqrt{2 p})}\right)^{\frac{1}{\beta}}.
\end{equation}
The FPT  distribution is finally deduced from $\sigma(s,n|1) $ and yields the following exact asymptotics (see SM):
\begin{equation}\label{eq : full_fpt_SATW}
F_{\underline{0}}(n|s_0=1) \underset{n \rightarrow \infty}{\sim} \frac{\Gamma(\frac{2}{\beta}-1)}{\Gamma(\frac{1}{2\beta}-\frac{1}{2})} 2^{-\frac{1+\beta}{2\beta}}n^{-\frac{1-\beta}{2\beta}-1}.
\end{equation}
While the $n$ decay is in agreement with the recent determination of the persistent exponent of the SATW relying on a different approach \cite{BarbierChebbah:2020aa}, this formalism based on the joint law gives access to the explicit expression of the prefactor for this strongly non Markovian process.  


{\it Conclusion.}  We have proposed   a general method to   derive explicit expressions of the  joint distribution  of   the first-passage time to a target and the number of distinct sites visited when the target is reached for $1d$ random walks. This method yields explicit expressions for several representative examples of Markovian search processes. Furthermore, we showed that the  dependence of the joint distribution on its space and time variables is captured by   a general scaling form, which  holds even for non Markovian processes. We argue that the joint distribution could have  applications in various situations where only partial  information -- either kinetic or geometric -- on trajectories is accessible ; in addition, it appears to be a useful technical tool that for instance can give  access to persistence properties of self interacting random walks.


\begin{thebibliography}{51}
\expandafter\ifx\csname natexlab\endcsname\relax\def\natexlab#1{#1}\fi
\expandafter\ifx\csname bibnamefont\endcsname\relax
  \def\bibnamefont#1{#1}\fi
\expandafter\ifx\csname bibfnamefont\endcsname\relax
  \def\bibfnamefont#1{#1}\fi
\expandafter\ifx\csname citenamefont\endcsname\relax
  \def\citenamefont#1{#1}\fi
\expandafter\ifx\csname url\endcsname\relax
  \def\url#1{\texttt{#1}}\fi
\expandafter\ifx\csname urlprefix\endcsname\relax\def\urlprefix{URL }\fi
\providecommand{\bibinfo}[2]{#2}
\providecommand{\eprint}[2][]{\url{#2}}

\bibitem[{\citenamefont{Hughes}(1995)}]{Hughes:1995}
\bibinfo{author}{\bibfnamefont{B.}~\bibnamefont{Hughes}},
  \emph{\bibinfo{title}{Random walks and random environments}}
  (\bibinfo{publisher}{New York: Oxford University Press},
  \bibinfo{year}{1995}).

\bibitem[{\citenamefont{B{\'e}nichou et~al.}(2011)\citenamefont{B{\'e}nichou,
  Loverdo, Moreau, and Voituriez}}]{Benichou:2011fk}
\bibinfo{author}{\bibfnamefont{O.}~\bibnamefont{B{\'e}nichou}},
  \bibinfo{author}{\bibfnamefont{C.}~\bibnamefont{Loverdo}},
  \bibinfo{author}{\bibfnamefont{M.}~\bibnamefont{Moreau}}, \bibnamefont{and}
  \bibinfo{author}{\bibfnamefont{R.}~\bibnamefont{Voituriez}},
  \bibinfo{journal}{Reviews of Modern Physics} \textbf{\bibinfo{volume}{83}},
  \bibinfo{pages}{81} (\bibinfo{year}{2011}).

\bibitem[{\citenamefont{Viswanathan et~al.}(2008)\citenamefont{Viswanathan,
  Raposo, and da~Luz}}]{Viswanathan:2008}
\bibinfo{author}{\bibfnamefont{G.~M.} \bibnamefont{Viswanathan}},
  \bibinfo{author}{\bibfnamefont{E.~P.} \bibnamefont{Raposo}},
  \bibnamefont{and} \bibinfo{author}{\bibfnamefont{M.~G.~E.}
  \bibnamefont{da~Luz}}, \bibinfo{journal}{Physics of Life Reviews}
  \textbf{\bibinfo{volume}{5}}, \bibinfo{pages}{133} (\bibinfo{year}{2008}).

\bibitem[{\citenamefont{Redner}(2001)}]{Redner:2001a}
\bibinfo{author}{\bibfnamefont{S.}~\bibnamefont{Redner}},
  \emph{\bibinfo{title}{A guide to First- Passage Processes}}
  (\bibinfo{publisher}{Cambridge University Press, Cambridge, England},
  \bibinfo{year}{2001}).

\bibitem[{\citenamefont{Metzler et~al.}(2014)\citenamefont{Metzler, Oshanin,
  and Redner}}]{bookSid2014}
\bibinfo{author}{\bibfnamefont{R.}~\bibnamefont{Metzler}},
  \bibinfo{author}{\bibfnamefont{G.}~\bibnamefont{Oshanin}}, \bibnamefont{and}
  \bibinfo{author}{\bibfnamefont{S.}~\bibnamefont{Redner}},
  \emph{\bibinfo{title}{First passage problems: recent advances}}
  (\bibinfo{publisher}{World Scientific, Singapore}, \bibinfo{year}{2014}).

\bibitem[{\citenamefont{Bray et~al.}(2013)\citenamefont{Bray, Majumdar, and
  Schehr}}]{Bray:2013}
\bibinfo{author}{\bibfnamefont{A.~J.} \bibnamefont{Bray}},
  \bibinfo{author}{\bibfnamefont{S.~N.} \bibnamefont{Majumdar}},
  \bibnamefont{and} \bibinfo{author}{\bibfnamefont{G.}~\bibnamefont{Schehr}},
  \bibinfo{journal}{Advances in Physics} \textbf{\bibinfo{volume}{62}},
  \bibinfo{pages}{225} (\bibinfo{year}{2013}).

\bibitem[{\citenamefont{Condamin et~al.}(2007)\citenamefont{Condamin,
  B{\'e}nichou, Tejedor, Voituriez, and Klafter}}]{Condamin2007}
\bibinfo{author}{\bibfnamefont{S.}~\bibnamefont{Condamin}},
  \bibinfo{author}{\bibfnamefont{O.}~\bibnamefont{B{\'e}nichou}},
  \bibinfo{author}{\bibfnamefont{V.}~\bibnamefont{Tejedor}},
  \bibinfo{author}{\bibfnamefont{R.}~\bibnamefont{Voituriez}},
  \bibnamefont{and} \bibinfo{author}{\bibfnamefont{J.}~\bibnamefont{Klafter}},
  \bibinfo{journal}{Nature} \textbf{\bibinfo{volume}{450}}, \bibinfo{pages}{77}
  (\bibinfo{year}{2007}).

\bibitem[{\citenamefont{B{\'e}nichou and Voituriez}(2014)}]{Benichou:2014fk}
\bibinfo{author}{\bibfnamefont{O.}~\bibnamefont{B{\'e}nichou}}
  \bibnamefont{and}
  \bibinfo{author}{\bibfnamefont{R.}~\bibnamefont{Voituriez}},
  \bibinfo{journal}{Physics Reports} \textbf{\bibinfo{volume}{539}},
  \bibinfo{pages}{225} (\bibinfo{year}{2014}).

\bibitem[{\citenamefont{Cheviakov et~al.}(2010)\citenamefont{Cheviakov, Ward,
  and Straube}}]{Cheviakov:2010}
\bibinfo{author}{\bibfnamefont{A.~F.} \bibnamefont{Cheviakov}},
  \bibinfo{author}{\bibfnamefont{M.~J.} \bibnamefont{Ward}}, \bibnamefont{and}
  \bibinfo{author}{\bibfnamefont{R.}~\bibnamefont{Straube}},
  \bibinfo{journal}{Multiscale Modeling \& Simulation}
  \textbf{\bibinfo{volume}{8}}, \bibinfo{pages}{836} (\bibinfo{year}{2010}).

\bibitem[{\citenamefont{Schuss et~al.}(2007)\citenamefont{Schuss, Singer, and
  Holcman}}]{Schuss2007}
\bibinfo{author}{\bibfnamefont{Z.}~\bibnamefont{Schuss}},
  \bibinfo{author}{\bibfnamefont{A.}~\bibnamefont{Singer}}, \bibnamefont{and}
  \bibinfo{author}{\bibfnamefont{D.}~\bibnamefont{Holcman}},
  \bibinfo{journal}{Proc Natl Acad Sci U S A} \textbf{\bibinfo{volume}{104}},
  \bibinfo{pages}{16098} (\bibinfo{year}{2007}).

\bibitem[{\citenamefont{Brummelhuis and Hilhorst}(1991)}]{Brummelhuis:1991ys}
\bibinfo{author}{\bibfnamefont{M.~J. A.~M.} \bibnamefont{Brummelhuis}}
  \bibnamefont{and} \bibinfo{author}{\bibfnamefont{H.~J.}
  \bibnamefont{Hilhorst}}, \bibinfo{journal}{Physica A: Statistical Mechanics
  and its Applications} \textbf{\bibinfo{volume}{176}}, \bibinfo{pages}{387}
  (\bibinfo{year}{1991}).

\bibitem[{\citenamefont{Brummelhuis and Hilhorst}(1992)}]{Brummelhuis:1992}
\bibinfo{author}{\bibfnamefont{M.~J. A.~M.} \bibnamefont{Brummelhuis}}
  \bibnamefont{and} \bibinfo{author}{\bibfnamefont{H.~J.}
  \bibnamefont{Hilhorst}}, \bibinfo{journal}{Physica A: Statistical Mechanics
  and its Applications} \textbf{\bibinfo{volume}{185}}, \bibinfo{pages}{35}
  (\bibinfo{year}{1992}).

\bibitem[{\citenamefont{Chupeau et~al.}(2015)\citenamefont{Chupeau, Benichou,
  and Voituriez}}]{Chupeau:2015sf}
\bibinfo{author}{\bibfnamefont{M.}~\bibnamefont{Chupeau}},
  \bibinfo{author}{\bibfnamefont{O.}~\bibnamefont{Benichou}}, \bibnamefont{and}
  \bibinfo{author}{\bibfnamefont{R.}~\bibnamefont{Voituriez}},
  \bibinfo{journal}{Nat Phys} \textbf{\bibinfo{volume}{11}},
  \bibinfo{pages}{844} (\bibinfo{year}{2015}).

\bibitem[{\citenamefont{B\'enichou et~al.}(2005)\citenamefont{B\'enichou,
  Coppey, Moreau, Suet, and Voituriez}}]{Benichou:2005a}
\bibinfo{author}{\bibfnamefont{O.}~\bibnamefont{B\'enichou}},
  \bibinfo{author}{\bibfnamefont{M.}~\bibnamefont{Coppey}},
  \bibinfo{author}{\bibfnamefont{M.}~\bibnamefont{Moreau}},
  \bibinfo{author}{\bibfnamefont{P.~H.} \bibnamefont{Suet}}, \bibnamefont{and}
  \bibinfo{author}{\bibfnamefont{R.}~\bibnamefont{Voituriez}},
  \bibinfo{journal}{EPL (Europhysics Letters)} \textbf{\bibinfo{volume}{70}},
  \bibinfo{pages}{42} (\bibinfo{year}{2005}).

\bibitem[{\citenamefont{Weiss and Calabrese}(1996)}]{Weiss:1996}
\bibinfo{author}{\bibfnamefont{G.~H.} \bibnamefont{Weiss}} \bibnamefont{and}
  \bibinfo{author}{\bibfnamefont{P.~P.} \bibnamefont{Calabrese}},
  \bibinfo{journal}{Physica A: Statistical and Theoretical Physics}
  \textbf{\bibinfo{volume}{234}}, \bibinfo{pages}{443} (\bibinfo{year}{1996}).

\bibitem[{\citenamefont{Burov and Barkai}(2007)}]{Burov:2007}
\bibinfo{author}{\bibfnamefont{S.}~\bibnamefont{Burov}} \bibnamefont{and}
  \bibinfo{author}{\bibfnamefont{E.}~\bibnamefont{Barkai}},
  \bibinfo{journal}{Physical Review Letters} \textbf{\bibinfo{volume}{98}},
  \bibinfo{pages}{250601} (\bibinfo{year}{2007}).

\bibitem[{\citenamefont{Condamin et~al.}(2008)\citenamefont{Condamin, Tejedor,
  Voituriez, Benichou, and Klafter}}]{Condamin:2008}
\bibinfo{author}{\bibfnamefont{S.}~\bibnamefont{Condamin}},
  \bibinfo{author}{\bibfnamefont{V.}~\bibnamefont{Tejedor}},
  \bibinfo{author}{\bibfnamefont{R.}~\bibnamefont{Voituriez}},
  \bibinfo{author}{\bibfnamefont{O.}~\bibnamefont{Benichou}}, \bibnamefont{and}
  \bibinfo{author}{\bibfnamefont{J.}~\bibnamefont{Klafter}},
  \bibinfo{journal}{Proceedings of the National Academy of Sciences}
  \textbf{\bibinfo{volume}{105}}, \bibinfo{pages}{5675} (\bibinfo{year}{2008}).

\bibitem[{\citenamefont{Weiss}(1994)}]{Weiss:1994}
\bibinfo{author}{\bibfnamefont{G.}~\bibnamefont{Weiss}},
  \emph{\bibinfo{title}{Aspects and Applications of the Random Walk}}
  (\bibinfo{publisher}{Amsterdam, Netherlands: North-Holland},
  \bibinfo{year}{1994}).

\bibitem[{\citenamefont{Larralde et~al.}(1992)\citenamefont{Larralde, Trunfio,
  Havlin, Stanley, and Weiss}}]{Larralde:1992a}
\bibinfo{author}{\bibfnamefont{H.}~\bibnamefont{Larralde}},
  \bibinfo{author}{\bibfnamefont{P.}~\bibnamefont{Trunfio}},
  \bibinfo{author}{\bibfnamefont{S.}~\bibnamefont{Havlin}},
  \bibinfo{author}{\bibfnamefont{H.~E.} \bibnamefont{Stanley}},
  \bibnamefont{and} \bibinfo{author}{\bibfnamefont{G.~H.} \bibnamefont{Weiss}},
  \bibinfo{journal}{Nature} \textbf{\bibinfo{volume}{355}},
  \bibinfo{pages}{423} (\bibinfo{year}{1992}).

\bibitem[{\citenamefont{Berezhkovskii et~al.}(1989)\citenamefont{Berezhkovskii,
  Makhnovskii, and Suris}}]{Berezhkovskii:1989}
\bibinfo{author}{\bibfnamefont{A.~M.} \bibnamefont{Berezhkovskii}},
  \bibinfo{author}{\bibfnamefont{Y.~A.} \bibnamefont{Makhnovskii}},
  \bibnamefont{and} \bibinfo{author}{\bibfnamefont{R.~A.} \bibnamefont{Suris}},
  \bibinfo{journal}{Journal of Statistical Physics}
  \textbf{\bibinfo{volume}{57}}, \bibinfo{pages}{333} (\bibinfo{year}{1989}).

\bibitem[{\citenamefont{Blumen et~al.}(1986)\citenamefont{Blumen, Klafter, and
  Zumofen}}]{Blumen:1986}
\bibinfo{author}{\bibfnamefont{A.}~\bibnamefont{Blumen}},
  \bibinfo{author}{\bibfnamefont{J.}~\bibnamefont{Klafter}}, \bibnamefont{and}
  \bibinfo{author}{\bibfnamefont{G.}~\bibnamefont{Zumofen}}, in
  \emph{\bibinfo{booktitle}{Optical Spectroscopy of Glasses}}, edited by
  \bibinfo{editor}{\bibfnamefont{I.}~\bibnamefont{Zschokke}}
  (\bibinfo{publisher}{Reidel Publ., Dordrecht}, \bibinfo{year}{1986}).

\bibitem[{\citenamefont{D.Ben-Avraham and S.Havlin}(2000)}]{D.Ben-Avraham:2000}
\bibinfo{author}{\bibnamefont{D.Ben-Avraham}} \bibnamefont{and}
  \bibinfo{author}{\bibnamefont{S.Havlin}}, \emph{\bibinfo{title}{Diffusion and
  reactions in fractals and disordered systems}} (\bibinfo{publisher}{Cambridge
  University Press}, \bibinfo{year}{2000}).

\bibitem[{\citenamefont{Dayan and Havlin}(1992)}]{Dayan:1992}
\bibinfo{author}{\bibfnamefont{I.}~\bibnamefont{Dayan}} \bibnamefont{and}
  \bibinfo{author}{\bibfnamefont{S.}~\bibnamefont{Havlin}},
  \bibinfo{journal}{Journal of Physics A: Mathematical and General}
  \textbf{\bibinfo{volume}{25}}, \bibinfo{pages}{L549} (\bibinfo{year}{1992}).

\bibitem[{\citenamefont{Kearney and Majumdar}(2005)}]{Kearney:2005wo}
\bibinfo{author}{\bibfnamefont{M.~J.} \bibnamefont{Kearney}} \bibnamefont{and}
  \bibinfo{author}{\bibfnamefont{S.~N.} \bibnamefont{Majumdar}},
  \textbf{\bibinfo{volume}{38}}, \bibinfo{pages}{4097} (\bibinfo{year}{2005}).

\bibitem[{\citenamefont{Krapivsky et~al.}(2010)\citenamefont{Krapivsky,
  Majumdar, and Rosso}}]{Krapivsky:2010wt}
\bibinfo{author}{\bibfnamefont{P.~L.} \bibnamefont{Krapivsky}},
  \bibinfo{author}{\bibfnamefont{S.~N.} \bibnamefont{Majumdar}},
  \bibnamefont{and} \bibinfo{author}{\bibfnamefont{A.}~\bibnamefont{Rosso}},
  \textbf{\bibinfo{volume}{43}}, \bibinfo{pages}{315001}
  (\bibinfo{year}{2010}).

\bibitem[{\citenamefont{Klinger et~al.}(2021)\citenamefont{Klinger, Voituriez,
  and B{\'e}nichou}}]{Klinger:2021aa}
\bibinfo{author}{\bibfnamefont{J.}~\bibnamefont{Klinger}},
  \bibinfo{author}{\bibfnamefont{R.}~\bibnamefont{Voituriez}},
  \bibnamefont{and}
  \bibinfo{author}{\bibfnamefont{O.}~\bibnamefont{B{\'e}nichou}},
  \bibinfo{journal}{Physical Review E} \textbf{\bibinfo{volume}{103}},
  \bibinfo{pages}{032107} (\bibinfo{year}{2021}).

\bibitem[{\citenamefont{Randon-Furling and
  Majumdar}(2007)}]{Randon-Furling:2007vu}
\bibinfo{author}{\bibfnamefont{J.}~\bibnamefont{Randon-Furling}}
  \bibnamefont{and} \bibinfo{author}{\bibfnamefont{S.~N.}
  \bibnamefont{Majumdar}}, \textbf{\bibinfo{volume}{2007}},
  \bibinfo{pages}{P10008} (\bibinfo{year}{2007}).

\bibitem[{\citenamefont{Evans and Majumdar}(2011)}]{Evans:2011}
\bibinfo{author}{\bibfnamefont{M.~R.} \bibnamefont{Evans}} \bibnamefont{and}
  \bibinfo{author}{\bibfnamefont{S.~N.} \bibnamefont{Majumdar}},
  \bibinfo{journal}{Physical Review Letters} \textbf{\bibinfo{volume}{106}},
  \bibinfo{pages}{160601} (\bibinfo{year}{2011}).

\bibitem[{\citenamefont{Evans et~al.}(2020)\citenamefont{Evans, Majumdar, and
  Schehr}}]{Evans:2020aa}
\bibinfo{author}{\bibfnamefont{M.~R.} \bibnamefont{Evans}},
  \bibinfo{author}{\bibfnamefont{S.~N.} \bibnamefont{Majumdar}},
  \bibnamefont{and} \bibinfo{author}{\bibfnamefont{G.}~\bibnamefont{Schehr}},
  \textbf{\bibinfo{volume}{53}}, \bibinfo{pages}{193001}
  (\bibinfo{year}{2020}).

\bibitem[{\citenamefont{Rosenstock}(1961)}]{Rosenstock:1961}
\bibinfo{author}{\bibfnamefont{H.~B.} \bibnamefont{Rosenstock}},
  \bibinfo{journal}{SIAM J. Appl. Math.} \textbf{\bibinfo{volume}{9}},
  \bibinfo{pages}{169} (\bibinfo{year}{1961}).

\bibitem[{\citenamefont{Amit et~al.}(1983)\citenamefont{Amit, Parisi, and
  Peliti}}]{amit_asymptotic_1983}
\bibinfo{author}{\bibfnamefont{D.~J.} \bibnamefont{Amit}},
  \bibinfo{author}{\bibfnamefont{G.}~\bibnamefont{Parisi}}, \bibnamefont{and}
  \bibinfo{author}{\bibfnamefont{L.}~\bibnamefont{Peliti}},
  \bibinfo{journal}{Physical Review B} \textbf{\bibinfo{volume}{27}},
  \bibinfo{pages}{1635} (\bibinfo{year}{1983}), \bibinfo{note}{publisher:
  American Physical Society}.

\bibitem[{\citenamefont{Pemantle}(2007)}]{pemantle_survey_2007}
\bibinfo{author}{\bibfnamefont{R.}~\bibnamefont{Pemantle}},
  \bibinfo{journal}{Probability Surveys} \textbf{\bibinfo{volume}{4}},
  \bibinfo{pages}{1} (\bibinfo{year}{2007}).

\bibitem[{\citenamefont{Grassberger}(2017)}]{grassberger_self-trapping_2017}
\bibinfo{author}{\bibfnamefont{P.}~\bibnamefont{Grassberger}},
  \bibinfo{journal}{Physical Review Letters} \textbf{\bibinfo{volume}{119}}
  (\bibinfo{year}{2017}).

\bibitem[{\citenamefont{Foster et~al.}(2009)\citenamefont{Foster, Grassberger,
  and Paczuski}}]{foster_reinforced_2009}
\bibinfo{author}{\bibfnamefont{J.~G.} \bibnamefont{Foster}},
  \bibinfo{author}{\bibfnamefont{P.}~\bibnamefont{Grassberger}},
  \bibnamefont{and} \bibinfo{author}{\bibfnamefont{M.}~\bibnamefont{Paczuski}},
  \bibinfo{journal}{New Journal of Physics} \textbf{\bibinfo{volume}{11}},
  \bibinfo{pages}{023009} (\bibinfo{year}{2009}), \bibinfo{note}{publisher: IOP
  Publishing}.

\bibitem[{\citenamefont{Stevens and Othmer}(1997)}]{stevens_aggregation_1997}
\bibinfo{author}{\bibfnamefont{A.}~\bibnamefont{Stevens}} \bibnamefont{and}
  \bibinfo{author}{\bibfnamefont{H.~G.} \bibnamefont{Othmer}},
  \bibinfo{journal}{SIAM Journal on Applied Mathematics}
  \textbf{\bibinfo{volume}{57}}, \bibinfo{pages}{1044} (\bibinfo{year}{1997}).

\bibitem[{\citenamefont{Sapozhnikov}(1994)}]{sapozhnikov_self-attracting_1994}
\bibinfo{author}{\bibfnamefont{V.~B.} \bibnamefont{Sapozhnikov}},
  \bibinfo{journal}{Journal of Physics A: Mathematical and General}
  \textbf{\bibinfo{volume}{27}}, \bibinfo{pages}{L151} (\bibinfo{year}{1994}).

\bibitem[{\citenamefont{Boyer and Walsh}(2010)}]{boyer_modelling_2010}
\bibinfo{author}{\bibfnamefont{D.}~\bibnamefont{Boyer}} \bibnamefont{and}
  \bibinfo{author}{\bibfnamefont{P.~D.} \bibnamefont{Walsh}},
  \bibinfo{journal}{Philosophical Transactions of the Royal Society A:
  Mathematical, Physical and Engineering Sciences}
  \textbf{\bibinfo{volume}{368}}, \bibinfo{pages}{5645} (\bibinfo{year}{2010}).

\bibitem[{\citenamefont{Boyer et~al.}(2012)\citenamefont{Boyer, Crofoot, and
  Walsh}}]{boyer_non-random_2012}
\bibinfo{author}{\bibfnamefont{D.}~\bibnamefont{Boyer}},
  \bibinfo{author}{\bibfnamefont{M.~C.} \bibnamefont{Crofoot}},
  \bibnamefont{and} \bibinfo{author}{\bibfnamefont{P.~D.} \bibnamefont{Walsh}},
  \bibinfo{journal}{Journal of The Royal Society Interface}
  \textbf{\bibinfo{volume}{9}}, \bibinfo{pages}{842} (\bibinfo{year}{2012}),
  \bibinfo{note}{publisher: Royal Society}.

\bibitem[{\citenamefont{B{\"o}rger et~al.}(2008)\citenamefont{B{\"o}rger,
  Dalziel, and Fryxell}}]{borger_are_2008}
\bibinfo{author}{\bibfnamefont{L.}~\bibnamefont{B{\"o}rger}},
  \bibinfo{author}{\bibfnamefont{B.~D.} \bibnamefont{Dalziel}},
  \bibnamefont{and} \bibinfo{author}{\bibfnamefont{J.~M.}
  \bibnamefont{Fryxell}}, \bibinfo{journal}{Ecology Letters}
  \textbf{\bibinfo{volume}{11}}, \bibinfo{pages}{637} (\bibinfo{year}{2008}).

\bibitem[{\citenamefont{Falc\'on-Cort\'es
  et~al.}(2017)\citenamefont{Falc\'on-Cort\'es, Boyer, Giuggioli, and
  Majumdar}}]{PhysRevLett.119.140603}
\bibinfo{author}{\bibfnamefont{A.}~\bibnamefont{Falc\'on-Cort\'es}},
  \bibinfo{author}{\bibfnamefont{D.}~\bibnamefont{Boyer}},
  \bibinfo{author}{\bibfnamefont{L.}~\bibnamefont{Giuggioli}},
  \bibnamefont{and} \bibinfo{author}{\bibfnamefont{S.~N.}
  \bibnamefont{Majumdar}}, \bibinfo{journal}{Phys. Rev. Lett.}
  \textbf{\bibinfo{volume}{119}}, \bibinfo{pages}{140603}
  (\bibinfo{year}{2017}).

\bibitem[{\citenamefont{d'Alessandro et~al.}(2021)\citenamefont{d'Alessandro,
  Barbier-Chebbah, Cellerin, B{\'e}nichou, M{\`e}ge, Voituriez, and
  Ladoux}}]{dAlessandro:2021vx}
\bibinfo{author}{\bibfnamefont{J.}~\bibnamefont{d'Alessandro}},
  \bibinfo{author}{\bibfnamefont{A.}~\bibnamefont{Barbier-Chebbah}},
  \bibinfo{author}{\bibfnamefont{V.}~\bibnamefont{Cellerin}},
  \bibinfo{author}{\bibfnamefont{O.}~\bibnamefont{B{\'e}nichou}},
  \bibinfo{author}{\bibfnamefont{R.-M.} \bibnamefont{M{\`e}ge}},
  \bibinfo{author}{\bibfnamefont{R.}~\bibnamefont{Voituriez}},
  \bibnamefont{and} \bibinfo{author}{\bibfnamefont{B.}~\bibnamefont{Ladoux}},
  \bibinfo{journal}{Nature Communications} 
  \textbf{\bibinfo{volume}{12}},\bibinfo{pages}{4118}
  (\bibinfo{year}{2021}).

\bibitem[{\citenamefont{Kampen}(1992)}]{Kampen:1992a}
\bibinfo{author}{\bibfnamefont{N.~V.} \bibnamefont{Kampen}},
  \emph{\bibinfo{title}{Stochastic Processes in Physics and Chemistry}}
  (\bibinfo{publisher}{North -Holland}, \bibinfo{year}{1992}).

\bibitem[{\citenamefont{Ernst}(1988)}]{Ernst:1988fk}
\bibinfo{author}{\bibfnamefont{M.~H.} \bibnamefont{Ernst}},
  \bibinfo{journal}{Journal of Statistical Physics}
  \textbf{\bibinfo{volume}{53}}, \bibinfo{pages}{191} (\bibinfo{year}{1988}).

\bibitem[{\citenamefont{Tejedor et~al.}(2012)\citenamefont{Tejedor, Voituriez,
  and B{\'e}nichou}}]{Tejedor:2012ly}
\bibinfo{author}{\bibfnamefont{V.}~\bibnamefont{Tejedor}},
  \bibinfo{author}{\bibfnamefont{R.}~\bibnamefont{Voituriez}},
  \bibnamefont{and}
  \bibinfo{author}{\bibfnamefont{O.}~\bibnamefont{B{\'e}nichou}},
  \bibinfo{journal}{Physical Review Letters} \textbf{\bibinfo{volume}{108}},
  \bibinfo{pages}{088103} (\bibinfo{year}{2012}).

\bibitem[{\citenamefont{Majumdar et~al.}(2015)\citenamefont{Majumdar,
  Sabhapandit, and Schehr}}]{PhysRevE.92.052126}
\bibinfo{author}{\bibfnamefont{S.~N.} \bibnamefont{Majumdar}},
  \bibinfo{author}{\bibfnamefont{S.}~\bibnamefont{Sabhapandit}},
  \bibnamefont{and} \bibinfo{author}{\bibfnamefont{G.}~\bibnamefont{Schehr}},
  \bibinfo{journal}{Phys. Rev. E} \textbf{\bibinfo{volume}{92}},
  \bibinfo{pages}{052126} (\bibinfo{year}{2015}).

\bibitem[{\citenamefont{Kusmierz et~al.}(2014)\citenamefont{Kusmierz, Majumdar,
  Sabhapandit, and Schehr}}]{PhysRevLett.113.220602}
\bibinfo{author}{\bibfnamefont{L.}~\bibnamefont{Kusmierz}},
  \bibinfo{author}{\bibfnamefont{S.~N.} \bibnamefont{Majumdar}},
  \bibinfo{author}{\bibfnamefont{S.}~\bibnamefont{Sabhapandit}},
  \bibnamefont{and} \bibinfo{author}{\bibfnamefont{G.}~\bibnamefont{Schehr}},
  \bibinfo{journal}{Phys. Rev. Lett.} \textbf{\bibinfo{volume}{113}},
  \bibinfo{pages}{220602} (\bibinfo{year}{2014}).

\bibitem[{\citenamefont{Majumdar et~al.}(2010)\citenamefont{Majumdar, Rosso,
  and Zoia}}]{Majumdar:2010}
\bibinfo{author}{\bibfnamefont{S.~N.} \bibnamefont{Majumdar}},
  \bibinfo{author}{\bibfnamefont{A.}~\bibnamefont{Rosso}}, \bibnamefont{and}
  \bibinfo{author}{\bibfnamefont{A.}~\bibnamefont{Zoia}},
  \bibinfo{journal}{Physical Review Letters} \textbf{\bibinfo{volume}{104}}
  (\bibinfo{year}{2010}).

\bibitem[{\citenamefont{Levernier et~al.}(2018)\citenamefont{Levernier,
  B{\'e}nichou, Gu{\'e}rin, and Voituriez}}]{Levernier:2018qf}
\bibinfo{author}{\bibfnamefont{N.}~\bibnamefont{Levernier}},
  \bibinfo{author}{\bibfnamefont{O.}~\bibnamefont{B{\'e}nichou}},
  \bibinfo{author}{\bibfnamefont{T.}~\bibnamefont{Gu{\'e}rin}},
  \bibnamefont{and}
  \bibinfo{author}{\bibfnamefont{R.}~\bibnamefont{Voituriez}},
  \bibinfo{journal}{Physical Review E} \textbf{\bibinfo{volume}{98}},
  \bibinfo{pages}{022125} (\bibinfo{year}{2018}).

\bibitem[{\citenamefont{MANDELBROT and VANNESS}(1968)}]{MANDELBROT1968}
\bibinfo{author}{\bibfnamefont{B.~B.} \bibnamefont{Mandelbrot}}
  \bibnamefont{and} \bibinfo{author}{\bibfnamefont{J.}~\bibnamefont{Vanness}},
  \bibinfo{journal}{Siam Review} \textbf{\bibinfo{volume}{10}},
  \bibinfo{pages}{422} (\bibinfo{year}{1968}).

\bibitem[{\citenamefont{Bicout and Burkhardt}(2000)}]{Bicout:2000}
\bibinfo{author}{\bibfnamefont{D.~J.} \bibnamefont{Bicout}} \bibnamefont{and}
  \bibinfo{author}{\bibfnamefont{T.~W.} \bibnamefont{Burkhardt}},
  \bibinfo{journal}{Journal of Physics A: Mathematical and General}
  \textbf{\bibinfo{volume}{33}}, \bibinfo{pages}{6835} (\bibinfo{year}{2000}).

\bibitem[{\citenamefont{Barbier-Chebbah
  et~al.}(2020)\citenamefont{Barbier-Chebbah, Benichou, and
  Voituriez}}]{BarbierChebbah:2020aa}
\bibinfo{author}{\bibfnamefont{A.}~\bibnamefont{Barbier-Chebbah}},
  \bibinfo{author}{\bibfnamefont{O.}~\bibnamefont{Benichou}}, \bibnamefont{and}
  \bibinfo{author}{\bibfnamefont{R.}~\bibnamefont{Voituriez}},
  \bibinfo{journal}{Physical Review E} \textbf{\bibinfo{volume}{102}},
  \bibinfo{pages}{062115} (\bibinfo{year}{2020}).

\end{thebibliography}

\end{document}